\documentstyle[prd,aps,epsf]{revtex}
\begin{document}
\title{\bf Phase Structure of 
Dynamical Triangulation Models in Three Dimensions } 
\author{\bf Ray L. Renken }
\address{ Department of Physics, University of Central Florida,
Orlando, Florida, 32816 }
\author{\bf Simon M. Catterall}
\address{ Department of Physics, Syracuse University,
Syracuse, New York, 13244 }
\author{\bf John B. Kogut }
\address{ Department of Physics, University of Illinois,
Urbana, Illinois, 61801 }
\date{ December 9, 1997 }
\maketitle
\mediumtext
\widetext
\begin{abstract}
\begin{quotation}

The dynamical triangulation model of three--dimensional quantum gravity is
shown to have a line of transitions in an expanded phase diagram which
includes a coupling $\mu$ to the order of the vertices. Monte Carlo
renormalization group and finite size scaling techniques are used to locate
and characterize this line. Our results indicate that for $\mu<\mu_1\sim -1.0$
the model is always in a {\it crumpled} phase independent of the value of the
curvature coupling.  For $\mu<0$ the results are in agreement with an
approximate mean field treatment.  We find evidence that this line corresponds
to {\it first order} transitions extending to positive $\mu$.  However the
behavior appears to change for $\mu>\mu_2\sim 2.0-4.0$. The simplest scenario
that is consistent with the data is the existence of a critical end point.  
\end{quotation}
\vskip .25in
\end{abstract}
\section{ Dynamical Triangulations with a Measure Term }
Triangulations provide a discretization of curved Euclidean spacetimes.  In
the Regge approach, a specific simplicial lattice is chosen a priori and the
properties of the spacetime are determined by the length of the lattice links.
A complementary approach, dynamical triangulations, fixes the length
of the links to an invariant cut-off and allows the lattice connectivity to
determine the geometry of the spacetime.  This paper uses the latter approach.
A lattice representation of the Einstein-Hilbert action is
\begin{equation}
S = \alpha N_0 - \beta N_D \label{action0}
\end{equation}
where $N_0$ is the number of vertices and $N_D$ the number
of simplices in the triangulation (the latter is then also the volume), D is the
dimension of the system and $\alpha$ and $\beta$ are corresponding chemical
potentials.  The coupling $\beta$ serves as the cosmological constant while
$\alpha$ is related to Newton's gravitational constant.  The connectivity
of a triangulation plays the same role as the metric in continuous manifolds
in the sense that points joined by a link are considered close together while
those not connected are considered further apart.
Continuum theories of quantum gravity generally involve
an integration over all possible physically inequivalent metrics.  In 
dynamical triangulation models, a sum over triangulations (i.e a sum over all
allowed connectivities) fulfills the same need.  Consequently, the partition
function for the dynamical triangulation model of quantum gravity is
\begin{equation}
Z = \sum_T \rho\left(T\right)  e^S \label{partition}
\end{equation}
where the sum is over all possible triangulations with fixed topology, $S$ is
the action given above and the term $\rho(T)$ allows for the possibility of a
non-trivial measure in the space of triangulations.  Matter can be coupled to
the gravity by adding appropriate terms to eqn. (\ref{action0}) and including
a sum over the matter degrees of freedom in eqn. (\ref{partition}).

These models of quantum gravity have proven themselves in two dimensions where
analytical results are available both in the continuum and on the lattice.
They agree in all cases where comparisons are possible.  This agreement occurs
with trivial measure $\rho(T)=1$.  In higher dimensions, it may be necessary
to utilize a non-trivial measure term in order to obtain an acceptable lattice
theory.  In this paper, the three--dimensional model is modified so as to
include such a measure term.  The resulting phase diagram is studied in detail.

There is additional motivation for adding a measure term.
As in lattice gauge theories, it is expected that
dynamical triangulation models 
must have a second order phase transition in their phase diagram if they are
to have a continuum limit and therefore a chance at physical relevance.
Many two--dimensional models have the necessary continuous transition.
However, in three dimensions,
while the simple action of eqn. (\ref{action0}) does have a phase transition,
it is first order.  The phase diagram is one--dimensional (parameterized by
$\alpha$) since $\beta$ is used to fix the volume.  Thus, it is
natural to expand the phase diagram by including some new coupling so that
there is a larger territory in which to search for continuous phase
transitions.  Expanded phase diagrams have been examined in the past, produced
by adding
both spin matter \cite{spin} and gauge matter \cite{gauge} to the system, but
no second order phase transition was found.  
Modifying the action $S$ to include a suitable measure term can be viewed as
another attempt to expand the phase diagram so as to produce a continuous
phase transition. The expectation is that the transition point $\alpha_t$ at
$\mu=0$ will extend into a transition line $\alpha_t\left(\mu\right)$ in
the $(\alpha,\mu)$ plane.

In \cite{Brug} a lattice version of a family of measures $\rho(T)$ is
introduced by adding a new term to the action
\begin{equation}
S = \alpha N_0 - \beta N_D + \mu M \label{action}
\end{equation}
where
\begin{equation}
M = \sum_{i \in N_0} \ln \left( { O_i \over {D + 1} } \right) 
\end{equation}
and $O_i$ is the number of $D$-simplices which include the vertex $i$, commonly
referred to as the ``order'' of $i$.  This additional term in the action
corresponds to a continuum measure of the form
\begin{equation}
\prod_x g^{\mu / 2}
\end{equation}
($g$ being the determinant of the metric).  In two dimensions, it is known from
Monte Carlo renormalization group \cite{thor} and exact calculations
\cite{kazakov} that any lattice theory with $\mu >0$ yields the same continuum
limit: this new term behaves as an irrelevant operator in the renormalization
group sense.  In higher dimensions, it is not known whether the operator
changes the fixed point structure of the theory and if so, how the bare $\mu$
should be chosen to approach a non-trivial continuum limit. These are the
questions we attempt to address here for the case $D=3$.

\section{ Computational Techniques }
We have used Monte Carlo simulations to do computations based on eqn. (3).
Three techniques have been most useful in analyzing the simulations and drawing
conclusions from them: the renormalization group approach, finite size
scaling, and histograms of the time series data.

The renormalization group has proven to be a powerful technique for
understanding field theory and statistical mechanics in flat space. When
combined with Monte Carlo simulation it allows a {\it non-perturbative} 
determination of the flows and phase structure.  A formulation of the
renormalization group applicable to dynamical triangulations has been developed
and established in previous papers\cite{rr1,thor,algo,rr2}.
It is useful to recall some details of this
scheme.  One iteration of the renormalization group transformation
eliminates a single vertex from the lattice.  This transformation is iterated
until the number of vertices in the blocked system is reduced to a target
number.  Two operators are used to monitor the flow in the expectation values:
$<N_3>$ and $<M>$. Their values after blocking depend both on
the initial volume of the system and the bare couplings. The latter may be used
to label different flow lines while the former parametrizes distance along such
a flow line.    For $\mu \le 0$, as the renormalization group transformation is
iterated, all flows initially head toward a common area in the $(N_3, M)$
plane.  Eventually, as the iterations continue, the flows diverge and head in
directions determined by the phase the couplings correspond to.  For large
positive $\mu$, the flows diverge immediately.

The renormalization group scheme has been tested previously (in three
dimensions) only near the transition $\alpha=\alpha_t\left(\mu\right)$
for $\mu = 0$.  It is important to have an independent means of checking
whether or not the renormalization group prediction of flows, transition
couplings, and transition orders is actually correct for $\mu\ne 0$.
If consistent, these independent means serve to
strengthen the case provided by the renormalization group.
Vertex susceptibilities $\chi$ provide just such a check.  

\begin{equation}
\chi = (<N_0^2> - <N_0>^2) / N_3
\end{equation}

\noindent
Close to a phase transition this quantity will typically exhibit a peak.
The position of this peak is called the pseudocritical coupling. In the
infinite volume limit this pseudocritical coupling converges to
the true transition coupling. The difference between this transition
coupling and the pseudocritical coupling is one measure of
the magnitude of finite size effects. The scaling of
the height of the peak in $\chi$ with system volume can be used to extract a
scaling exponent $\omega$. A value for $\omega$ of unity would
indicate a first order transition. Smaller values are
consistent with continuous critical behavior.

If a first order transition is strong enough, it can be seen in
the Monte Carlo time series.  A double peak structure in a histogram of the
vertex number data is a classic sign of a first order phase
transition. Continuous transitions possess only a single peak in
this histogram.

For fixed values of $\mu$, we search for a transition as a function
of $\alpha$ using the above techniques.  A variety of values of $\mu$ are
considered, including both positive and negative values.  
All runs are for (quasi) fixed volumes $N_3$ in the range of $4000$ to
$16000$. Typically ten million sweeps were performed where one sweep is
defined as $N_3$ attempted elementary moves.  For values of $\mu\sim 4$,
ten million was found to be much too small a number of sweeps.

\section{ Results for the Transition Line }
\subsection{ $\mu=0$.} 
\begin{figure}
\centering
\leavevmode
\epsfxsize=4.0in \epsfbox{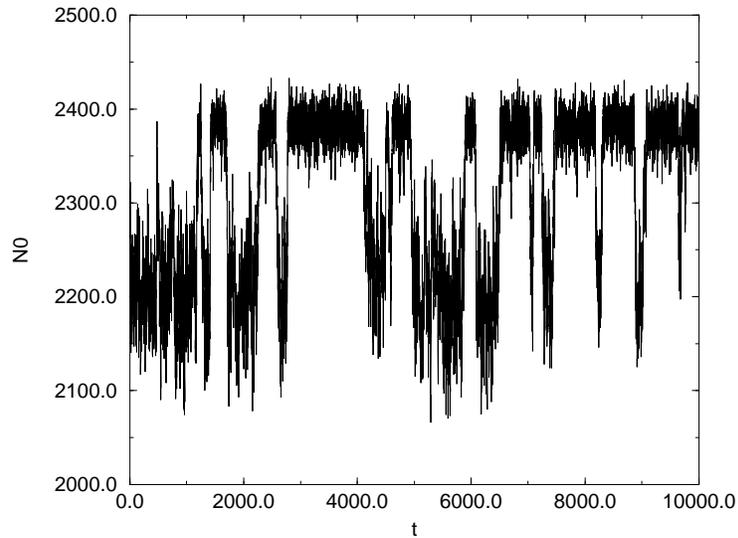}
\caption{Time history of the number of vertices during a simulation at
$\mu = 0$, $\alpha = 4.0$, and $N_3 = 8000$.  Time is in units of $1000 N_3$ 
attempted updates. }
\label{thist1}
\end{figure}

\begin{figure}
\centering
\leavevmode
\epsfxsize=4.0in \epsfbox{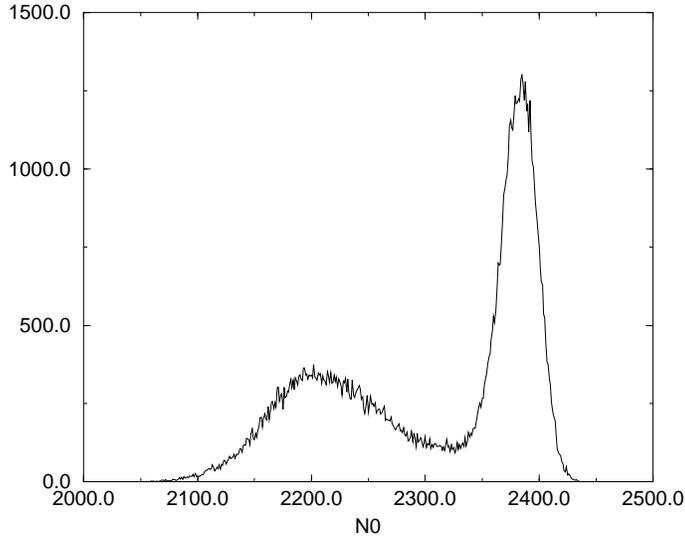}
\caption{A histogram of the data presented in fig. \ref{thist1}. }
\label{hist1}
\end{figure}

The transition
is known to be strongly first order for $\mu = 0$.  This is illustrated in
fig. \ref{thist1} which shows a Monte Carlo time series for the vertex number
$N_0$ and in fig. \ref{hist1} which displays a histogram of the same data.
The plots show clearly the existence of two metastable states connected
by tunneling events typical of a discontinuous phase transition. This
observation is in agreement with earlier studies \cite{3dearly}.
On one side of the transition the system is in a crumpled phase and on the
other side it is in a branched polymer phase.  The crumpled phase is
characterized by a vertex density $<N_0>/N_3$ which goes to zero as the volume
$N_3$ gets large.  In the branched polymer phase this ratio approaches $1/3$.

\begin{figure}
\centering
\leavevmode
\epsfxsize=4.0in \epsfbox{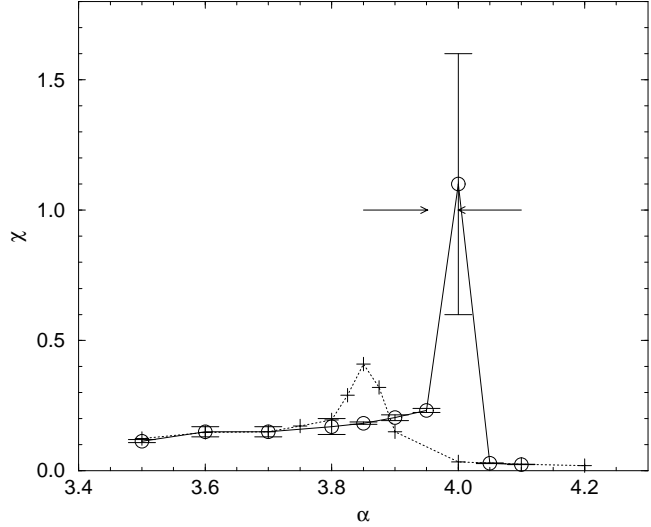}
\caption{ Vertex susceptibility as a function of $\alpha$ for $\mu = 0.0$.
Arrows indicate the renormalization group prediction for the transition region.}
\label{susc0}
\end{figure}

The vertex susceptibility also reveals the first order nature of
the transition:
in fig.\ref{susc0} results from a system with $8000$ simplices (plotted as
circles) clearly display a huge peak.  The error bar at the peak is so
large because there are relatively infrequent tunnelings between metastable
states at the transition.  Notice that there is a roughly constant
characteristic value of the susceptibility in the crumpled phase (small
$\alpha$) and a much smaller roughly constant characteristic value in the
branched polymer phase (large $\alpha$).  These features will be visible at
other values of $\mu$ as well. The arrows denote the range of
critical coupling determined from a study of the renormalization group flows
(computed previously \cite{rr1} and not reproduced here).
Notice that
it is consistent with the position of the peak in $\chi$.  This is an
indication that finite size effects for $N_3=8000$ are small at $\mu=0$.

However data from a system with $4000$ simplices
(with plus as the plotting symbol) is included in the plot for comparison.
(The error bars are roughly the size of the plotting symbols and were left off
for aesthetic purposes).  The peak is not nearly as high for the smaller
system.  For significantly smaller systems the peak would become
indistinguishable from the background.  Furthermore, it is clear that
the pseudocritical coupling is very far from
the true critical coupling for this lattice size implying that finite
size effects are large. The moral to be drawn from
this is that it would be very hard to infer
the transition order correctly from lattices which are too small.

\subsection{ $\mu=-0.50$.}
\begin{figure}
\centering
\leavevmode
\epsfxsize=4.0in \epsfbox{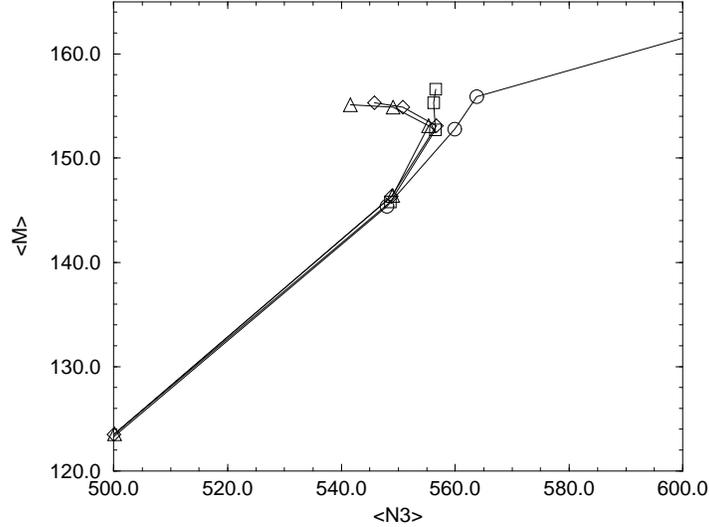}
\caption{Renormalization group flows with $\mu = -0.50$. }
\label{mu50}
\end{figure}

For $\mu \ne 0$ the transition value of $\alpha$ has typically not been
previously determined (however, see recent work in \cite{new_Hot}).  The
renormalization group flows provide a convenient
method of searching for the transition since the flows eventual diverge,
flowing toward small node density in the crumpled phase and large node density
in the branched polymer phase.  Fig. \ref{mu50} illustrates this with several
values of $\alpha$
chosen to be near the transition.  Flows similar to the case of $\mu=0$ are
seen here.  Each line represents a different value of $\alpha$ and motion along
that line (up and to the right) represents the evolution of the volume $N_3$
and the average of $M$ as the initial
volume is increased.  Since the renormalization group is iterated until the
number of vertices in the blocked system is the same as the number of vertices
in the original system when the original volume is $500$, systems with larger
initial volumes require more vertex deletions.  Hence, motion along the line
corresponds to expectation values for effective theories at larger and
larger length scales.  The values of $\alpha$ are, reading from right to left
in the figure, 4.7, 4.8, 4.9, and 5.0.  The rightmost line, which continues
out of the boundaries of the plot, moves toward larger volumes.  Since the
vertex number is fixed this means the vertex density is heading toward zero.
This is characteristic of the crumpled phase.  The leftmost lines cannot go
left indefinitely because there is a lower bound on the volume given a
number of vertices.  These leftmost flows correspond to the branched polymer
phase.  One flow ends up moving neither right nor left.  The coupling for
this flow is presumably near the transition value.  Further points along the
flow (i.e. larger volumes) are needed to determine which phase it is in or
whether it approaches a fixed point.  

At a strongly first order transition,
such as in three dimensions with $\mu = 0$, it is not necessary to go to
large volumes to see what the phase is, so there are no intermediate flows
\cite{rr2}.  In four dimensions, the corresponding transition appears second
order at small volumes, and there are intermediate flows similar to the one
in the figure.  At larger volumes, there is now evidence that this latter
transition is actually first order \cite{4dtransition}. Thus the 
three--dimensional models at (moderate) negative $\mu$ seem to behave somewhat
similarly to the the four--dimensional theory at $\mu=0$. We might expect them
to display large finite size effects and this is verified below:
it is necessary to go to large volumes to get
the pseudocritical couplings determined from vertex
susceptibilities to agree with the renormalization group predictions.

\begin{figure}
\centering
\leavevmode
\epsfxsize=4.0in \epsfbox{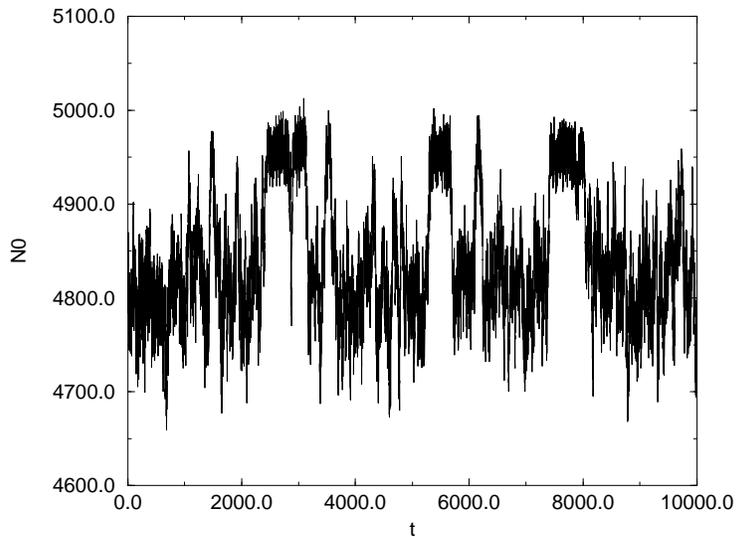}
\caption{Time history of the number of vertices during a simulation at
$\mu = -0.50$, $\alpha = 4.85$, and $N_3 = 16000$. Time is in units of
$1000 N_3$ attempted updates. }
\label{thist2}
\end{figure}

\begin{figure}
\centering
\leavevmode
\epsfxsize=4.0in \epsfbox{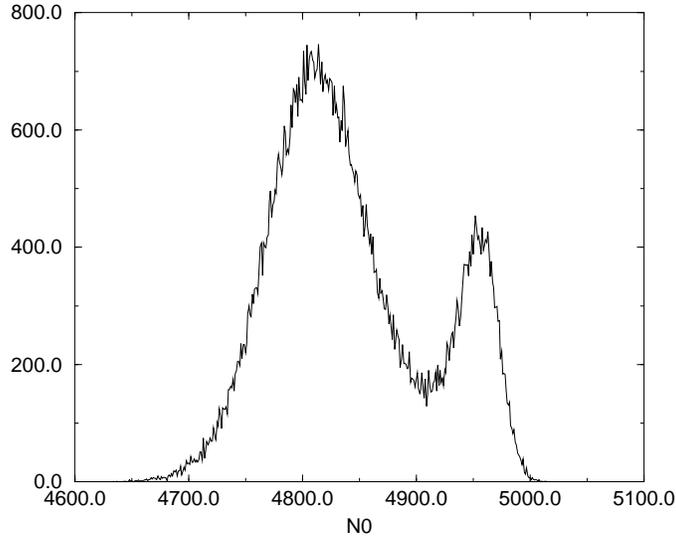}
\caption{A histogram of the data presented in fig. \ref{thist2}. }
\label{hist2}
\end{figure}

Fig. \ref{thist2} shows the time history for a simulation at the transition
and fig. \ref{hist2} is a histogram of that data.
The existence of two metastable states as revealed in these plots lends
support to the idea that the transition remains first order as $\mu$ is
varied towards negative values.

\begin{figure}
\centering
\leavevmode
\epsfxsize=4.0in \epsfbox{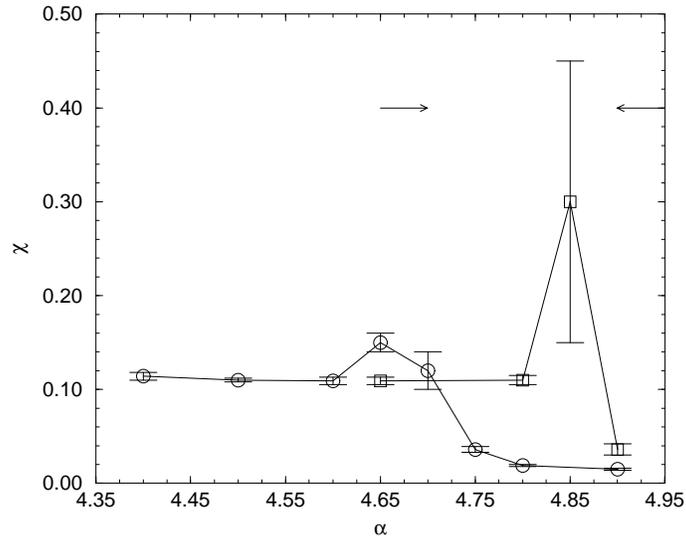}
\caption{Vertex susceptibility as a function of $\alpha$ for $\mu = -0.50$.
Arrows indicate the renormalization group prediction for the transition region.}
\label{susc50}
\end{figure}

The vertex susceptibility is shown in fig. \ref{susc50}
There are plateaus of characteristic values in both the crumpled and the
branched polymer phases.  This time, for $N_3 = 8000$ (with circles as the
plotting symbol), the plateaus are separated by a much smaller peak.  This
peak is noticeably to the left of the range of transition values suggested
by the renormalization group flows (obtained with maximum volume $N_3=8000$).
Data for $N_3 = 16000$ (with squares as
the plotting symbol) show that the transition is moving toward greater
$\alpha$ as the volume increases.  This is more consistent with the
renormalization group result.  This large change in the pseudocritical
coupling is indicative of the presence of large finite volume corrections.
The large error bar on the peak for $N_3 = 16000$ is again due to the
presence of just a few fluctuations between metastable states.

\subsection{ $\mu=-0.75$.}
For the limited lattice sizes used in this study the situation becomes somewhat
unclear at large negative coupling. For moderate lattice volumes there are no
clear signals of metastability and a histogram of the vertex number shows no
strong double peak structure.  One might argue that the explanation is that the
latent heat is small in this regime -- perhaps going to zero for some critical
$\mu$ \cite{new_Hot}.  However, as we have shown our data
indicate that finite size effects get
progressively larger for increasing negative $\mu$. These finite
size effects can obscure the true critical behavior -- on small to moderate
lattice sizes the first order transition can appear continuous or even absent. 
Indeed if we assume the correct order parameter for the transition
is $<N_0/N_3>$ and if we assume this vanishes in the crumpled phase
(supported by the mean field arguments we present later) then the observed
latent heat is {\it increasing} as $\mu$ is made more negative. This would
seem to argue against a critical end point for $\mu<0$.

\begin{figure}
\centering
\leavevmode
\epsfxsize=4.0in \epsfbox{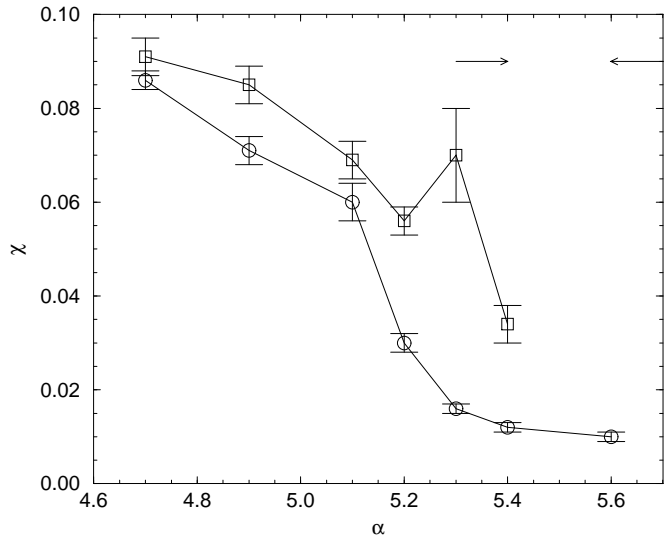}
\caption{ Vertex susceptibility as a function of $\alpha$ for $\mu = -0.75$.
Arrows indicate the renormalization group prediction for the transition region.}
\label{susc75}
\end{figure}

Further evidence supporting this scenario is presented in 
fig. \ref{susc75} which shows the
vertex susceptibility for $\mu=-0.75$. For $8000$ simplices
(circles), the peak has completely vanished
and is replaced with a gradual interpolation between the two characteristic
values.  The range of transition values indicated by the
renormalization group is now even further to the right of the crossover region
in the susceptibilities which is an indication that
finite size effects are even larger here than for $\mu=-0.50$. 
The data for $16000$ simplices (squares) shows the
beginnings of a peak closer to the critical regime predicted by the
renormalization group method (which used a maximum of 8000 simplices).  It
appears that finite size effects dominate the physics even at this large volume
and an even larger volume would be necessary to extract 
unambiguous results. We have not attempted such a simulation.

\subsection{ $\mu=2.0$.}
\begin{figure}
\centering
\leavevmode
\epsfxsize=4.0in \epsfbox{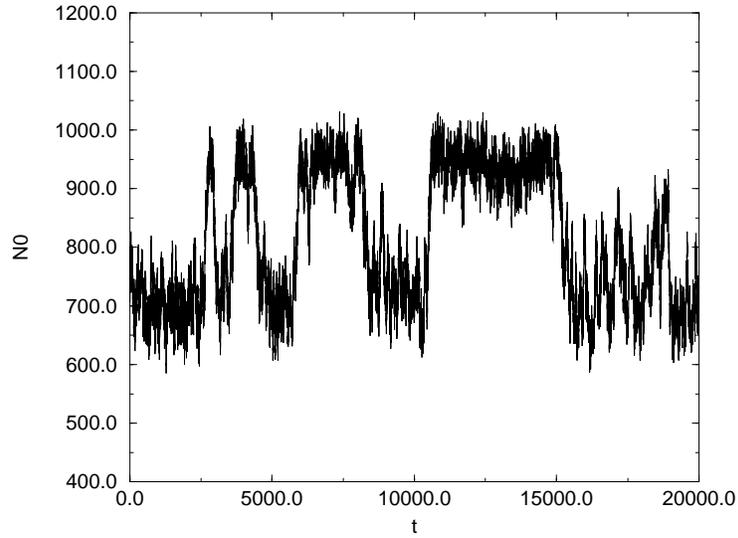}
\caption{Time history of the number of vertices during a simulation at
$\mu = 2$ and $N_3 = 4000$.}
\label{mu2thist}
\end{figure}

\begin{figure}
\centering
\leavevmode
\epsfxsize=4.0in \epsfbox{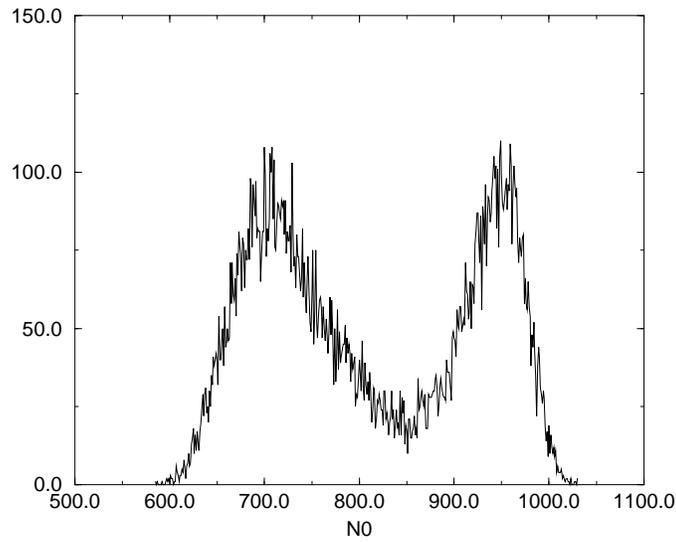}
\caption{A histogram of the data presented in fig. \ref{mu2thist}. }
\label{mu2hist}
\end{figure}

We have argued that the data favor a first order line for $\mu<0$. Our
evidence concerning the $\mu$ dependence of finite size effects 
suggests that simulations for $\mu>0$ may suffer {\it smaller}
finite size effects. This is indeed seen: 
for $1.0 < \mu < 2.0$ it becomes possible to see tunneling between metastable
states on smaller lattices ($N_3 = 4000$).  A time history
for $\mu=2$ and $N_3 = 4000$ is given in fig. \ref{mu2thist} and the
corresponding histogram is given in fig. \ref{mu2hist}.  Similar results
are obtained at $\mu = 1$.
Analysis of the vertex susceptibility and renormalization group flows also show
strong support for a first order phase transition at these
points on the phase boundary.

\subsection{ $\mu=4.0$.}
\begin{figure}
\centering
\leavevmode
\epsfxsize=4.0in \epsfbox{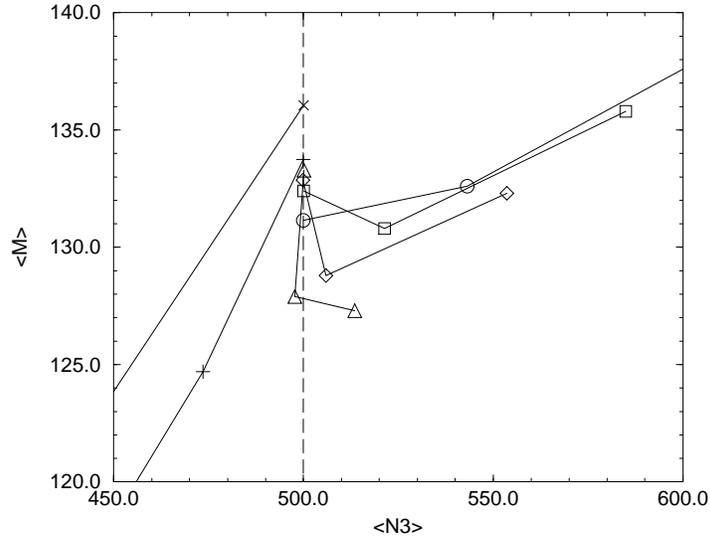}
\caption{Renormalization group flows with $\mu = 4.0$. }
\label{mup40}
\end{figure}

For $\mu = 4$, results are quite different.
The renormalization group flows are shown in fig.\ref{mup40}.  A dashed line
has been included in the plot as a reminder that the flows all start at
$N_3 = 500$.  Here the flows diverge immediately rather than briefly traveling
identical paths as they did for $\mu \le 0$.  The flows fan out more uniformly
across the plane.

\begin{figure}
\centering
\leavevmode
\epsfxsize=4.0in \epsfbox{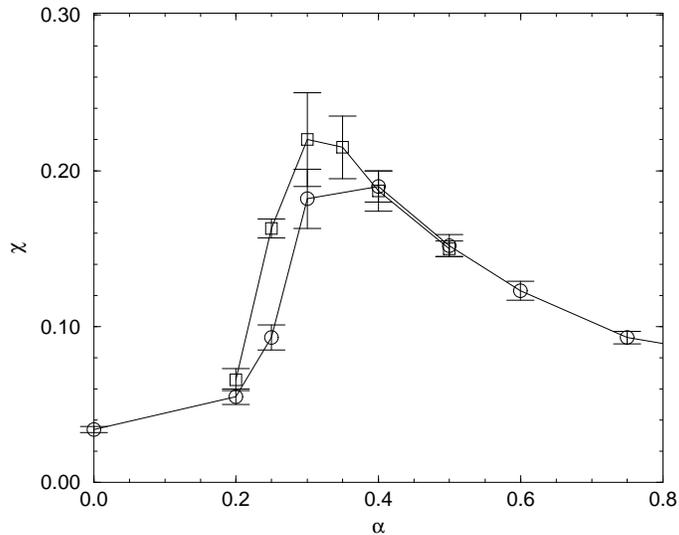}
\caption{Vertex susceptibility for $\mu=4.0$ }
\label{suscp40}
\end{figure}

Examining the vertex susceptibility (fig. \ref{suscp40}) reveals that there is
still a peak but that it fails to scale with volume.  There are no signs of
metastability for any value of $\alpha$.  There are also no indications
of large finite size effects, so the data suggest that there is no transition
at this value of $\mu$.  Perhaps the first order line terminates for a value
of $\mu$ between $2$ and $4$.  One other significant feature of the simulations
is that there the observed autocorrelation times become {\it extremely} long.
These autocorrelation times could be due to the presence of a nearby critical
end point.  It is also possible that for this value $\mu$ we have
entered a new phase which is
exceptionally difficult to simulate.

\subsection{ Summary }
\begin{figure}
\centering
\leavevmode
\epsfxsize=4.0in \epsfbox{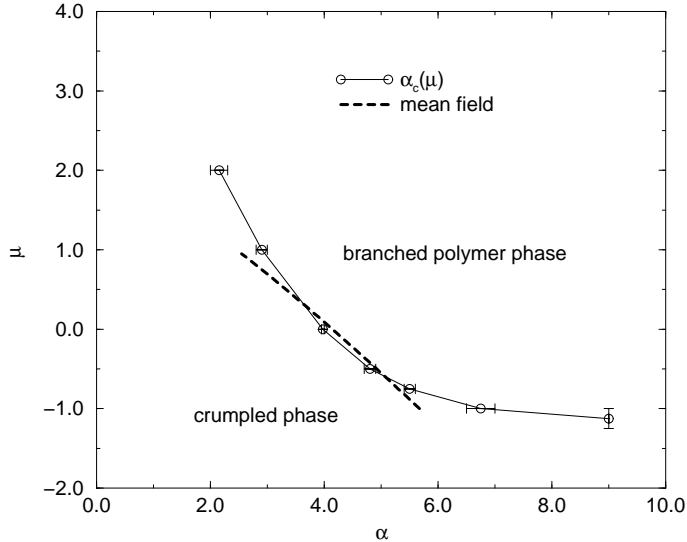}
\caption{ Phase diagram of three--dimensional triangulations with measure term.}
\label{trans}
\end{figure}

The above results as well as some data for other values of $\mu$ provide the
basis for a plot of the transition through the $(\alpha,\mu)$ plane.  The
results are displayed in fig. \ref{trans}.
The error bars in the
figure change direction at $\alpha = 9$ because for smaller values of $\alpha$
it is convenient to fix $\mu$ and vary $\alpha$ when searching for the phase
boundary while for larger values of $\alpha$ it is more convenient to fix
$\alpha$ and vary $\mu$. Notice that the transition appears to head off
to large values of $\alpha$ as $\mu$ is decreased further than $\mu\sim -1$.
Indeed our data is consistent with a curve which asymptotes to
a horizontal line.
Thus it appears that there is a value of $\mu$ below which there is no phase
transition as a function of $\alpha$: there is only a crumpled phase. 
This is because the transition line becomes horizontal.  
Our results suggest that the transition is first order for $\mu <= 2$ with
the first order line ending for $\mu$ somewhere between $2$ and $4$.
The dotted curve is the prediction of our mean
field treatment which will be discussed in the following section.

\section{Mean Field Arguments}

In a previous paper \cite{sing} we have discussed how the two phases of 
the dynamical triangulation models can be understood in terms of a condensation
of {\it singular} vertices. In general the crumpled phase in $D$ dimensions
consists of a single $(D-3)$-simplex common to a number of $D$-simplices which
diverges as a power of the total volume. As the vertex coupling $\alpha$ is
increased this structure becomes unstable disappearing entirely in the branched
polymer phase. The transition is driven by the fluctuations in this singular
structure \cite{fluc_sing}.  

With this as a motivation we have considered an ansatz for the partition
function which replaces the counting of distinct triangulations with the
problem of enumerating the possible orders of the vertices.

\begin{equation}
Z\left(N_0,N_3\right) =\sum_{n_i}\prod_{i=1}^{N_0}
\omega\left(n_i\right) \delta\left(\sum_{i=1}^{N_0}n_i-cN_3\right)
\end{equation}

The factors $\omega\left(n_i\right)$ give the probability that the vertex $i$
has order $n_i$ and the final $\delta$-function constraint reflects the fact
that the total order is proportional to the total volume \footnote{ Naively
$c=4$ in $D=3$.} 
Notice that this approach
is based on two approximations; that the counting of states can be done by
counting vertex orders and that the latter can be done in a mean field manner
by assuming that each varies independently of the others {\it except} for
the global constraint. Such a model describes well certain branched polymer
models \cite{bialas} and has been proposed as relevant to the description of
dynamical triangulations \cite{mother}.  This general {\it constrained mean
field} model was solved in \cite{cmf} for general probabilities
$\omega\left(n_i\right)$.  As discussed in \cite{sing}, for
describing dynamical triangulations the factors of
$\omega\left(n_i\right)$ should be taken to be the number of triangulations
of the $2$-sphere (in three dimensions) surrounding a given vertex $i$. 
We take the form
\cite{xxx}

\begin{equation}
\omega\left(n_i\right)=
\left(\frac{n_i}{4}\right)^{-7/2}\exp{\beta_2^c\left(n_i-4\right)}
\end{equation}

This has the correct asymptotics ($\beta_2^c=\ln{\frac{16}{3\sqrt{3}}}$) 
and yields unity for the minimal triangulation of the $2$-sphere with $4$
simplices. Notice that it is now trivial to incorporate the effects of the
measure term -- the power is simply modified from $-7/2$ to $-7/2+\mu$.
Using the results of \cite{cmf} one can write the partition function of
this model in the thermodynamic limit as 

\begin{equation}
Z=\frac{1}{2\pi}\exp{\left(\left(-\log{\lambda}+\beta_2^c\right)cN_3 + 
                          \ln{F_\mu\left(\lambda\right)}N_0\right) }
\label{cmf_eq}
\end{equation}

\noindent
where the parameter $\lambda$ is the solution of the equation ($\rho=N_0/N_3$)

\begin{equation}
\frac{c}{\rho}
=\frac{\lambda F^\prime_\mu\left(\lambda\right)}{F_\mu\left(\lambda\right)}
\end{equation}

\noindent
and, here 

\begin{equation}
F_\mu\left(\lambda\right)=
\frac{4^{7/2}}{\exp{4\beta_2^c}}\sum_{q=2}^{\infty}\left(2q\right)
^{-7/2+\mu}\lambda^{2q}.
\end{equation}

This model exhibits two phases; for 
$\rho< \rho_c=c\frac{F\left(1\right)}{F^\prime\left(1\right)}$
the parameter $\lambda=1$ and the system is in a collapsed phase where vertices
of small order behave independently and the global constraint is satisfied by a
small number of vertices which are shared by a number of simplices
on the order of the volume. This regime is
identified with the crumpled phase of the dynamical triangulation model. 
Conversely for $\rho> \rho_c$ the parameter $\lambda$ varies between
zero and one and the free energy
varies continuously with $\rho$. The distribution of vertex orders then behaves
as

\begin{equation}
p\left(n\right)\sim \omega\left(n\right)e^{-n\log{\lambda} }
\end{equation}

This phase is then identified with the branched polymer phase in the dynamical
triangulation model. As $\rho$ (and hence $N_0$) is varied (by varying the
coupling $\alpha$) the system moves between these two phases. In practice a
canonical ensemble in which the vertex number can fluctuate is used.

\begin{equation}
Z_C=\sum_{N_0}e^{\alpha N_0}Z\left(N_0,N_3\right)
\end{equation}

The phase diagram then contains two phases corresponding to
the crumpled and branched polymer phase separated by a {\it first order}
phase
transition \cite{new_bialas}. In the branched polymer
phase $N_0/N_3$ varies continuously with $\alpha$ tending
to $\rho_c$ as $\alpha-\alpha_t\left(\mu\right)\to 0^+$.
Conversely, $N_0/N_3\to 0$ in the crumpled phase
$\alpha<\alpha_t\left(\mu\right)$.
The phase boundary is predicted to be

\begin{equation}
\alpha_t\left(\mu\right)=-\log{F_\mu\left(1\right)}
\end{equation}

This curve is plotted against the numerically determined phase
boundary in fig. \ref{trans} for $1>\mu>-1$. The agreement is rather
impressive - at $\mu=0$ the predicted critical coupling is
$\alpha_t=4.1$ which is rather close to the best numerical
determinations $\alpha_t\sim 3.95-4.00$. It appears for low
density (the crumpled phase) the mean field approximation captures
much of the important physics.  The agreement becomes markedly
worse for $\mu>1$ -- indeed the mean field model would predict that
the system was always branched polymer for $\mu>3/2$. This
would seem to imply that the transition line would once again
become horizontal requiring large negative values of $\alpha$ to
enter the crumpled phase as $\mu\to \frac{3}{2}$. 
We do not see this in the simulations -- 
although the critical behavior of the model certainly seems to
undergo some form of change for sufficiently large positive $\mu$.

Finally, the mean field phase diagram offers the possibility
of understanding the single phase observed at $\mu<\mu_1\sim -1$. The
critical density $\rho_c$ increases with increasing negative $\mu$
approaching $\rho_c=1$ for $\mu\to -\infty$. However the
vertex density cannot increase beyond $\rho_{\rm max}=\frac{1}{3}$ -- so if
$\rho_c>\rho_{\rm max}=\frac{1}{3}$ the system will always be in a
crumpled phase which agrees qualitatively with
our simulations.

\section{ Conclusion }

Our studies of the three--dimensional dynamical triangulation model with
measure term allow
us to draw several conclusions. Firstly, the usual transition point of the model
can be extended into a transition line in the $(\alpha,\mu)$ plane. This line
separates a crumpled phase from a branched polymer phase.
This
boundary corresponds to a line of first order transitions.  We
infer the existence of this line using several methods: renormalization
group studies, traditional finite size scaling and an
approximate mean field treatment. All are consistent. 
It appears that finite volume effects become increasingly
important for negative $\mu$. This can make it
somewhat problematic to identify a first order transition. Indeed the
$D=3$ simulations for $\mu<0$ resemble in this respect the $D=4$
simulations at $\mu=0$. The renormalization group method
has shown itself to be extremely useful in
this context giving {\it much} better predictions for such
quantities as the infinite volume critical coupling than simple
finite size scaling. 

The critical behavior for positive $\mu$ appears to undergo
a change in the interval $\mu=2.0-4.0$. At $\mu=2.0$ there
appears to be a weak first order transition but there is no
sign of such a phase transition for $\mu=4.0$. Nevertheless 
there are very long autocorrelation times there. One simple 
interpretation of these observations is that the line of 
first order transitions has terminated on a (nearby) critical
end point. This might offer the
possibility of a continuum limit for the model. 
Other scenarios are possible too.  It is possible
that the critical line bifurcates into two and a new phase
appears.  Extensive numerical work will be required to
distinguish unambiguously between these possibilities.

\acknowledgements
This work was supported in part by NSF Grant PHY-9503371 and DOE
Grant DE-FG02-85ER40237.

\end{document}